\documentclass[12pt]{article}
\usepackage{epsf}
\def\beq{\begin{equation}}
\def\eeq{\end{equation}}
\def\ap#1#2#3 {Ann. Phys. (NY) {\bf#1} (19#2) #3}
\def\err#1#2#3 {{\it Erratum} {\bf#1} (19#2) #3}
\def\ib#1#2#3 {{\it ibid.} {\bf#1} (19#2) #3}
\def\ijmp#1#2#3 {Int. J. Mod. Phys. {\bf#1} (19#2) #3}
\def\jetp#1#2#3 {JETP Lett. {\bf#1} (19#2) #3}
\def\mpl#1#2#3 {Mod. Phys. Lett. {\bf#1} (19#2) #3}
\def\np#1#2#3 {Nucl. Phys. {\bf#1} (19#2) #3}
\def\pl#1#2#3 {Phys. Lett. {\bf#1} (19#2) #3}
\def\prep#1#2#3 {Phys. Rep. {\bf#1} (19#2) #3}
\def\prev#1#2#3 {Phys. Rev. {\bf#1} (19#2) #3}
\def\prl#1#2#3 {Phys. Rev. Lett. {\bf#1} (19#2) #3}
\def\sjnp#1#2#3 {Sov. J. Nucl. Phys. {\bf#1} (19#2) #3}
\def\spj#1#2#3 {Sov. Phys. JETP {\bf#1} (19#2) #3}
\def\spu#1#2#3 {Sov. Phys. Usp. {\bf#1} (19#2) #3}
\def\zp#1#2#3 {Zeit. Phys. {\bf#1} (19#2) #3}

\begin{document}
\begin{titlepage}
\begin{center}
{\Large \bf William I. Fine Theoretical Physics Institute \\
University of Minnesota \\}  \end{center}
\vspace{0.2in}
\begin{flushright}
FTPI-MINN-04/26 \\
UMN-TH-2314-04 \\
June 2004 \\
\end{flushright}
\vspace{0.3in}
\begin{center}
{\Large \bf  The rate of metastable vacuum decay in (2+1) dimensions
\\}
\vspace{0.2in}
{\bf M.B. Voloshin  \\ }
William I. Fine Theoretical Physics Institute, University of
Minnesota,\\ Minneapolis, MN 55455 \\
and \\
Institute of Theoretical and Experimental Physics, Moscow, 117259
\\[0.2in]
\end{center}

\begin{abstract}

The pre-exponential factor in the probability of decay of a metastable
vacuum is calculated for a generic (2+1) dimensional model in the limit
of small difference $\epsilon$ of the energy density between the
metastable and the stable vacua. It is shown that this factor is
proportional to $\epsilon^{-7/3}$ and that the power does not depend on
details of the underlying field theory. The calculation is done by using
the effective Lagrangian method for the relevant soft (Goldstone)
degrees of freedom in the problem. Unlike in the (1+1) dimensional case,
where the decay rate is completely determined by the parameters of the
effective Lagrangian and is thus insensitive to the specific details of
the underlying (microscopic) theory, in the considered here (2+1)
dimensional case the pre-exponential factor is found up to a constant,
which does depend on specifics of the underlying short-distance
dynamics, but does not depend on the energy asymmetry parameter
$\epsilon$. Thus the functional dependence of the decay rate on
$\epsilon$ is universally determined in the considered limit of small
$\epsilon$.

\end{abstract}

\end{titlepage}

The problem of decay of a metastable vacuum state has attracted interest
since long ago both in statistical physics\cite{ls,langer1,langer2,pp}
and in the relativistic field theory\cite{lw,vko,coleman,cc}. In the
latter setting the metastable (false) vacuum is a state of one or more
fields corresponding to a local rather than the global minimum of
the energy density. Such state is stable under small quantum
fluctuations of the field(s), however it can decay into a lower-energy
vacuum state through large fluctuations described by the quantum
tunneling. The mechanism of such transition\cite{vko,coleman} is quite
similar to that in the first order phase transitions and is described by
nucleation and subsequent expansion of bubbles of the lower (true)
vacuum amidst the bulk of the metastable phase. In the process of the
expansion the excess $\epsilon$ of the energy density in the metastable
vacuum (latent heat) is being converted into the energy of the
(expanding) surface of the bubble. Clearly, the classical
expansion of the bubble is possible only if the gain in the volume
energy $\epsilon \cdot (Volume)$ compensates for the positive energy of
the surface of the bubble $\mu \cdot (Surface ~ Area)$ with $\mu$
standing for the surface tension. The minimal configuration where these
are exactly equal is the critical spherical bubble with the radius
$R_c=(d-1) \mu/\epsilon$, where $d$ is the total number of the space and
time dimensions. Starting from $R_c$ the bubbles expand classically. The
evolution in the classically forbidden domain at $R < R_c$ is described
by the quantum tunneling.

It should be noted that in the above description the thickness of the
surface is totally ignored in comparison with the radius of the bubble.
The scale for the thickness of the transition region between the phases
is set by the Compton wavelengths of the relevant particles in either of
the vacua. Assuming that no relevant massless particles in either of the
vacua are present in this problem, one readily concludes that the thin
wall approximation is always valid in the limit of small $\epsilon$
which is assumed throughout the present paper.

The rate $\Gamma$ of nucleation of a critical bubble per unit time and
per unit volume is determined by  the tunneling part of its trajectory
in the classically forbidden domain, and the exponential factor, found
from the action on this trajectory, is well known\cite{vko,coleman} in
terms of $\mu$ and $\epsilon$:
\beq
\Gamma= F \, e^{-B}~,
\label{gam}
\eeq
where $B=V_d \, (d-1)^{d-1} \, \mu^d/ \epsilon^{d-1}$ with
$V_d= \pi^{d/2}/\Gamma(1+d/2)$ being the volume of a unit ball in $d$
dimensions. A calculation of the pre-exponential factor $F$ requires a
summation over the fluctuations of the field(s) near the tunneling
trajectory\cite{cc}. In a (1+1) dimensional model, i.e. for $d=2$, such
summation can be done entirely within the thin wall approximation, and
the factor $F$ in eq.(\ref{gam}) is given\cite{mv} by $F=\epsilon/(2
\pi)$. The latter expression is universal in the sense that the only
quantities that determine the rate $\Gamma$ are $\mu$ and $\epsilon$. In
other words, any possible complexity of the underlying field dynamics
reduces for the purpose of calculation of the rate $\Gamma$ to two
macroscopic  parameters\footnote{It can be also mentioned that the
resulting equation (\ref{gam}) in (1+1) dimensions has no corrections in
powers of the dimensionless parameter $\epsilon/\mu^2$. Only terms with
higher exponents of $- \pi \mu^2/\epsilon$ are possible\cite{mv}.} $\mu$
and $\epsilon$. The (1+1) dimensional expression for $F$ fully agrees
with calculations in specific field models\cite{stone,ks,mr}.

The purpose of the present paper is to consider the problem of
calculating the pre-exponential factor $F$ in a (2+1) dimensional
theory, i.e. for $d=3$. Unlike in the (1+1) dimensional case, the
knowledge of only the macroscopic parameters $\mu$ and $\epsilon$ is
insufficient for a complete calculation of $F$ for $d=3$. However it
will be shown that the dependence of the factor $F$ on the parameter
$\epsilon$ in the limit $\epsilon \to 0$ is still universal, while the
dependence on (a combination of) the parameter $\mu$ and any mass scales
in the underlying field theory, does depend on the details of the model
(i.e. on the `microscopic' dynamics).  Thus some universality still
remains in the (2+1) dimensional case, although in a substantially
reduced form. Namely, it will be shown here that in (2+1) dimensions the
rate of the false vacuum decay behaves at $\epsilon \to 0$ as
\beq
\Gamma={{\cal A} \over \epsilon^{7/3}} \, \exp \left ( - {16 \pi \over
3} \, {\mu^3 \over \epsilon^2} \right)~,
\label{res}
\eeq
where the dimensional constant ${\cal A}$
depends on the details, such as mass parameters and coupling constants,
of the underlying field model, but does not depend on $\epsilon$. The
parameter $\mu$ in eq.(\ref{res}) is the renormalized surface tension of
the boundary separating the two vacuum phases in the limit $\epsilon \to
0$. It can be noted that the power of $\epsilon$ in the pre-exponent in
eq.(\ref{res}) is in agreement with the result of a direct
calculation\cite{msr} in a specific $\phi^4$ model. The claim in the
present paper is that this power is universal in the limit $\epsilon \to
0$ so that it does not depend on details of specific model.

The calculation of the tunneling exponential factor as well as of the
pre-exponent is conveniently done in terms of the Euclidean-space
formulation of the model\cite{coleman,cc}. The partition function $Z$
for the metastable vacuum state is written in terms of the Euclidean
action $S$ in the standard form of the path integral:
\beq
Z={\cal N} \, \int e^{-S[\phi]} \, {\cal D} \phi
\label{pf}
\eeq
where ${\cal N}$ is the normalization factor and the path integration
runs over all the fields in the model, generically denoted here as
$\phi$, with the boundary condition that the fields approach their
values in the metastable vacuum at the boundaries of the normalization
space-time box. The metastability of the considered `vacuum' state
results in that the partition function develops imaginary part (similar
to the imaginary part of the energy of a resonance) which determines the
decay rate (per unit space-time volume) according to the
relation\cite{stone,coleman} $\Gamma = 2 {\rm Im} (\ln Z)/(VT)$
with $VT$ being the space-time volume of the normalization box.

The action functional $S[\phi]$ has a saddle point configuration
corresponding to the tunneling trajectory of the bubble, the so-called
bounce\cite{coleman}. In the thin wall approximation the bounce is a
three-dimensional ball of the `true' vacuum separated from the bulk of
the false vacuum by the thin wall with the surface tension $\mu$. The
radius $R_c$ of the bounce is thus determined by the extremum of the
effective action
\beq
S_{eff}=\mu \, A_B - \epsilon \, V_B~,
\label{seff}
\eeq
where $V_B$ is the volume of the ball and $A_B$ is its surface area.
Clearly, the extremum of the action is achieved at $R_c=2 \mu/\epsilon$,
which is the critical bubble size in (2+1) dimensions, and the value of
the action on the bounce is $S_B=B=(16 \pi/3) \, \mu^3/\epsilon^2$,
which gives the exponential factor in eq.(\ref{res}).

In order to calculate the pre-exponential factor one has to evaluate the
path integral $Z_1$ around the bounce configuration. The one-loop
expression for the rate then reads as\cite{cc}
\beq
\Gamma={1 \over VT} \left | {{\rm det}(S_1^{(2)}) \over {\rm
det}(S_0^{(2)})} \right |^{-1/2}\, e^{-B}~,
\label{grat}
\eeq
where $S_1^{(2)}$ is the operator of the second variation of the
(Euclidean) action at the bounce configuration, and $S_0^{(2)}$ is the
same operator for variation of the fields around the `flat' false vacuum
state. The operator $S_1^{(2)}$ has three translational zero modes and
the integration over those cancels against the space-time normalization
factor in the denominator. In Ref.\cite{cc} this integration has been
done explicitly. However in the present discussion it is somewhat more
convenient to leave the expression in the symbolic form (\ref{grat}) and
to deal with the zero modes later. Also it should be mentioned that the
operator $S_1^{(2)}$ has one negative mode, integration over which
produces the imaginary part of the partition function, as explained in
Ref.\cite{cc}.

The spectrum of the operator $S_1^{(2)}$ consists of two substantially
different parts: the hard part with the eigenvalues $\lambda_n$ starting
at the scale set by the mass parameters of the underlying field model,
which scale is generically denoted here as $m$, and the soft part, whose
scale is set by the (inverse) radius of the bounce. The soft part of the
spectrum is universal and does not depend on details of the underlying
field model as long as the condition $m R_c \gg 1$ is satisfied, which
is always the case in the thin wall limit, i.e. at $\epsilon \to 0$. The
explicit expression for the soft eigenvalues is\cite{cc}
\beq
\lambda_\ell={\ell (\ell+1)-2 \over R_c^2}~,
\label{soft}
\eeq
where $\ell=0, 1, 2, \ldots$ is the angular momentum and, obviously, the
degeneracy of each mode is $2 \ell+1$. It can be readily noticed that
the soft eigenmodes coincide (up to an overall normalization) with those
found from the effective action (\ref{seff}). Indeed, using the
parametrization of the surface of the bounce in polar coordinates as
$r(\theta, \varphi)$ and expanding the radial position $r$ around its
value at the extremum of the action: $r(\theta, \varphi)= R_c +
\sigma(\theta, \varphi)$, one gets in the quadratic order in the
deviation $\sigma$ the expression
\beq
S_{eff}=S_B+ {\mu \over 2} \, \int \left ( \partial_\alpha \sigma
\partial^\alpha \sigma - 2 \sigma^2 \right ) \,  d\Omega~,
\label{expan}
\eeq
where $\partial_\alpha \sigma \partial^\alpha \sigma = (\partial_\theta
\sigma)^2+ (\partial_\varphi \sigma)^2/\sin^2 \theta$ and $d\Omega =
\sin \theta \, d\theta \, d\varphi$. Clearly, the spectrum of the
quadratic part in eq.(\ref{expan}) is proportional to that in
eq.(\ref{soft}).

The described separation between the hard and the soft modes becomes
ambiguous at $\ell \sim m R_c$, where the soft part merges into the hard
one. The details of this merger would be unimportant if the path
integral over the soft spectrum were convergent. Then the whole
calculation would be reduced (by the Appelquist-Carazzone
theorem\cite{ac}) to calculating the path integral with the effective
action in eq.(\ref{seff}), i.e. in the effective `low-energy' theory.
This however is the case only in (1+1) dimensions\cite{mv}, while in the
discussed here (2+1) dimensional case the path integral with the
effective action (\ref{seff}) diverges, and one has to resort to a
somewhat more accurate application of the Appelquist-Carazzone theorem,
which requires\cite{ac} an explicit consideration of the regularization
of the effective low-energy theory.

In order to describe the regularization of the effective low-energy
theory we concentrate now on the notion of the parameter $\mu$ in the
limiting case of $\epsilon =0$. At zero $\epsilon$ the two considered
vacua are degenerate and there is a stable field configuration
interpolating between them. In the three dimensional Euclidean space the
interface between the vacua makes a two-dimensional surface with the
action proportional to the area of the surface:
\beq
S_{eff}=\mu \cdot Area~.
\label{seff2}
\eeq
The presence of the wall spontaneously breaks the translational
invariance. As a result there appears a spectrum of (Goldstone) modes,
that can be described by a massless two-dimensional scalar propagating
on the surface of the wall. The spectrum can also be readily found from
the effective action (\ref{seff2}) by considering small deviations of
the position of the wall from its equilibrium (flat) shape. On a flat
wall the spectrum of these modes can be parametrized by the
two-dimensional momentum $k_\alpha$ ($\alpha=1,2$): $\lambda_k = k^2$. A
calculation of the partition function in the effective `low-energy'
theory at the one-loop level  immediately runs into the problem that
the integral over the Goldstone modes is divergent. In particular, the
renormalization of $\mu$ in the effective theory is quadratically
divergent:
\beq
\mu \to \mu+{1 \over 2} \int \ln k^2 \, {d^2k \over (2\pi)^2}~.
\label{diverg}
\eeq
In the full theory (as opposed to the effective one) however, no such
divergence arises, since the effective description fails at $k \sim m$,
where the soft spectrum merges into the hard one. Thus physically the
ultraviolet cutoff in the integral in eq.(\ref{diverg}) is at the scale
$m$, and the surface tension gets a one-loop quantum correction of order
$m^2$, which is the normal behavior in the full theory. In order to
still enable a description of the low-energy modes by the effective
action (\ref{seff2}) at the loop level, it is necessary to explicitly
introduce a separation parameter, which would serve as an ultraviolet
regulator for the low-energy theory, while still being within the
applicability of the expression for the soft modes. This can be done by
using the standard Pauli-Villars regulator fields.

Following the Pauli-Villars procedure we introduce a set of regulator
fields $\psi_i$ (at least two are required to regularize the quadratic
divergence in eq.(\ref{diverg})) with the mass parameters $M_i$. For
each regulator field the spectrum of the eigenvalues is shifted up with
respect to those of the fields in the original field model ($\lambda_n$)
by $M_i^2$:
\beq
\lambda_n(\psi_i)=\lambda_n+M_i^2~.
\label{shift}
\eeq
The loop with the regulator field $\psi_i$ is given the weight factor
$c_i$ subject to the condition:
\beq
\sum_i c_i=1~,~~~~~ \sum_i c_i \, M_i^2=0~.
\label{rcond}
\eeq   
The regulator mass parameters $M_i$ are assumed to be much less than the
full theory mass scale: $M_i \ll m$, but much larger than the inverse
size of the surface of the wall: $M_i \gg (Area)^{-1/2}$.

Formally, the described regulator fields are introduced in the path
integral $Z$ of the original theory by inserting a factor of one in the
form:
\beq
Z={Z_\psi \over Z_\psi} \, Z~,
\label{zpsi}
\eeq
where $Z_\psi=\prod_i (Z[\psi_i])^{c_i}$. The partition functions can
then be split (at least at the one-loop level) into the products of the
soft ($Z_s$) and hard ($Z_h$) contributions, e.g. $Z=Z_s \, Z_h$, where
the separation between the ``soft" and ``hard" modes is introduced at a
scale $\Lambda$, intermediate between $M_i$ and $m$: $M_i \ll \Lambda
\ll m$. Then the identity (\ref{zpsi}) for the partition function can be
rewritten as
\beq
Z={Z_s \over (Z_\psi)_s} \, \left \{ (Z_\psi)_s \, Z_h \right \}~.
\label{separ}
\eeq
Clearly, the first factor is the regularized partition function
described by the effective action in eq.(\ref{seff2}). In particular the
regularized (at one loop) surface tension in this effective theory reads
as
\beq
\mu_{reg}=\mu+{1 \over 2} \, \int \left [ \ln k^2 - \sum_i c_i \ln
(k^2+M_i^2) \right ] \, {d^2k \over (2 \pi)^2}=\mu + {{\bar M}^2 \over 8
\pi}~,
\label{mur}
\eeq
where ${\bar M}^2=\sum_i c_i \, M_i^2 \, \ln M_i^2$.

The expression in the curly braces in eq.(\ref{separ}) is the original
path integral with the soft modes replaced by those of the regulator
fields, which implies that all the modes relevant for calculation of the
latter expression are hard in the sense that they start at least from
the scale of regulator masses $M_i$. Thus at any shape of the wall the
latter expression is sensitive only to local properties of the surface,
i.e. to higher curvatures. In particular, if the wall is curved with a
large radius $R$ this part can produce corrections to the effective
action in eq.(\ref{seff2}) of at most the relative magnitude $O(M^{-2}
R^{-2})$. In particular at the curvature corresponding to the radius
$R_c$ of the bounce such ``finite wall thickness"
corrections\footnote{An expression for these corrections in a $\phi^4$
model can be found in Ref.\cite{msr}.} are not singular in $\epsilon$ at
$\epsilon \to 0$. For this reason the discussed consideration of the
regularization procedure for a flat wall (at $\epsilon = 0$) is also
applicable at the intended level of accuracy for      a spherical wall
of the bounce at a small but finite $\epsilon$. In the latter case the
effective low-energy action (\ref{seff2}) can be replaced by the one in
eq.(\ref{seff}) since the term with $\epsilon$ requires no
regularization in the effective low-energy theory.

Thus the discussed problem of calculating the rate of the false vacuum
decay is reduced to evaluating the contribution of the bounce
configuration and of the fluctuations around it to the partition
function determined by the effective action (\ref{seff}) and regularized
by the `soft' regulator factor $(Z_\psi)_s$. As already mentioned the
value of the effective action (\ref{seff}) on the saddle point
configuration reproduces the exponential factor in the decay rate, while
the pre-exponential factor can be written as
\beq
F = f_0 \exp {1 \over 2} \, \sum_{\ell=0}^\infty (2 \ell +1)\, \left \{
\sum_i  c_i    \ln \left [ \ell (\ell+1) + M_i^2 R_c^2 -\omega^2 \right
]  -  \ln \left [ \ell (\ell+1)  -\omega^2 \right ]\right \}~,
\label{summ}
\eeq
where the constant $f_0$ comes from the expression in curly braces in
eq.(\ref{separ}) and does not depend on $\epsilon$ (or, equivalently, on
$R_c$), and the
parameter $\omega$ is temporarily introduced in order to regularize the
infrared singularity at $\omega^2=2$ due to the translational zero
modes. One can also notice that, formally, the upper limit in the sums
over $\ell$ shout be set at a large value $L$ related to the previously
introduced separation parameter $\Lambda$ as $L \sim \Lambda R_c$.
However the overall sum in eq.(\ref{summ}) is convergent (at $\ell_{max}
\sim M_i$) due to the Pauli-Villars constraints (\ref{rcond}), and the
sum can be extended to infinity as shown. Furthermore, the overall
normalization factor in the eigenvalues is not important, since the
total number of modes is the same for the regulator fields and the
original soft part of the spectrum, so that any common additive term
cancels in the total sum.

The sums in eq.(\ref{summ}) can be readily evaluated (up to a numerical
additive constant) using the Euler-Maclaurin summation formula with the
result reading (at $\omega^2$ close to 2) as
\beq
F = {\tilde f}_0 \, {(M R_c)^{\omega^2+1/3} \over
|2-\omega^2|^{3/2}} \, \exp \left ( - {1 \over 2} {\bar M}^2 \, R_c^2
\right) ~,
\label{sumd}
\eeq
where $\ln M = \sum_i c_i \, \ln M_i$, and  the constant ${\tilde f}_0$
differs from $f_0$ in eq.(\ref{summ}) only by a numerical factor. One
readily recognizes the term proportional to ${\bar M}^2$ in the
exponent as the renormalization, according to eq.(\ref{mur}) of the
surface
tension $\mu$ in the effective action (\ref{seff}), and this term,
together with the contribution of the hard modes (the expression in the
curly brackets in eq.(\ref{separ})) replaces the lowest-order surface
tension $\mu$ by the one with the one-loop correction in the leading
semiclassical exponent.

It is the factor with a power of $ M R_c$ in eq.(\ref{sumd}) which
produces a nontrivial power dependence of the pre-exponential factor $F$
on the parameter $\epsilon$, through the relation $R_c \propto
\epsilon^{-1}$. Although one can safely set $\omega^2=2$ in this term,
the dependence on $\omega$ is shown in eq.(\ref{sumd}) in order to
illustrate the origin of the contributions to this term: the $\omega^2$
part comes from the shift of the eigenvalues with respect to $\ell (\ell
+1)$, while the extra 1/3 in the power is due to the discretization of
the modes on a sphere. As will be discussed few lines below, a proper
treatment of the denominator in eq.(\ref{sumd}) singular at $\omega^2=2$
produces no extra dependence of the factor $F$ on $\epsilon$. Thus after
setting $\omega^2=2$ in the power of $M R_c$ one arrives at the
formula (\ref{res}) for the  $\epsilon^{-7/3}$ behavior of the
pre-exponential factor in the decay rate.

The singular at $\omega^2=2$ behavior arises in eq.(\ref{sumd}) due to
the space-time translational invariance in the probability of bubble
nucleation. This singularity can be readily dealt with by either the
standard consideration\cite{cc} of the integration over the
translational zero modes, or by using the following simple
regularization procedure in terms of the effective theory\cite{mv}. Let
us temporarily slightly break the translational invariance by
introducing
a dependence of the nucleation probability on the position $\vec x$ of
the center of the bounce described by the Gaussian factor $\exp(-\xi \,
x^2)$ with a small parameter $\xi$. The total probability of the
nucleation in a large space-time volume is then finite and is given by
\beq
\int \Gamma \, e^{-\xi x^2} \, d^3x= \left ( {\pi \over \xi} \right
)^{3/2} \, \Gamma~.
\label{gtot}
\eeq
On the other hand the shift of the center of the bounce is equivalent to
an amplitude of the partial wave with $\ell=1$ of the excitation
$\sigma$ of the surface of the bounce in eq.(\ref{expan}). Thus the
effect of the introduced infrared regularization is equivalent to adding
to the action (\ref{expan}) additional term $(3 \xi/4\pi) \int \sigma^2
\, d\Omega$. Although, formally, this term should be added only for the
partial wave with $\ell=1$ where it lifts the translational modes from
zero, it can be safely added to all modes, since for all other modes the
limit $\xi \to 0$ is nonsingular. As a result the described infrared
regularization is equivalent to the shift of the `frequency' from
$\omega^2=2$ to $\omega^2=2-3\xi/(2 \pi \, \mu)$. Using this regularized
expression in eq.(\ref{sumd}) one readily finds from comparison with
eq.(\ref{gtot}) that the rate $\Gamma$ (per unit space-time volume) is
finite in the limit $\xi \to 0$, and no new dependence on $\epsilon$ of
the factor $F$ is introduced by the infrared regularization (only an
extra dependence on $\mu$ does arise).

One can see from the presented here calculation that the pre-exponential
factor $\epsilon^{-7/3}$ in the rate of false vacuum decay in (2+1)
dimensions arises in fact as an analog of the Casimir effect on the
finite surface of the bounce due to the massless spectrum of excitations
of the surface waves (Goldstone modes) described by the effective action
(\ref{seff}). As previously mentioned, in a (1+1) dimensional case this
soft spectrum completely dominates and fully determines the relevant
path integral for calculating the rate. In the considered here (2+1)
dimensional case the importance of the soft part of the spectrum is
greatly weakened in comparison, but is still sufficient for deriving the
dependence on the large-scale parameter $R_c \propto \epsilon^{-1}$ of
the pre-exponential factor. It further looks highly unlikely that in a
(3+1) dimensional case the soft part of the spectrum alone can be used
to make any conclusions about the behavior of pre-exponential factor in
the rate of the false vacuum decay.

The presented here consideration, and the resulting formula in
eq.(\ref{res}) can be used essentially without modification for
the case of thermal decay of a metastable state in a first order phase
transition in a three-dimensional system (i.e. in (3+1) dimensions) near
the stability point, i.e. at $\epsilon \to 0$. However in realistic
thermal systems the nucleation of the bubbles is usually governed by a
diffusion dynamics, rather than by the conservative Hamiltonian one
implied throughout the discussion in this paper. The pre-exponential
factor in the nucleation rate in a diffusion dynamics case is a well
known textbook material\cite{lp} going back to the original work of
Zel'dovich\cite{yab}.

I thank L. Okun and A. Vainshtein for helpful discussions. I also
acknowledge the support by the Alexander von Humboldt Foundation of my
visit to the University of Bonn where this paper was finalized. This
work is supported in part by the DOE grant DE-FG02-94ER40823.

{\it Note added.} After this paper was completed I became aware of the
work \cite{garriga}, where the behavior of the pre-exponential factor
equivalent to $\epsilon^{-7/3}$ in (2+1) dimensions has been derived in
the context of membrane creation by an antisymmetric tensor field. I
believe that the calculation in the present paper is somewhat simpler,
and is more directly related to the generic framework of  one-loop
calculations in specific field-theoretical models.

\end{document}